\documentclass[pre,aps,showpacs,showkeys]{revtex4}
\usepackage{amsfonts}
\usepackage{amsmath}

\input{epsf}
\begin{document}
\title{A quantum dynamical framework for Brownian heat engines}
\author{G. S. Agarwal} 
\affiliation{Department of Physics, Oklahoma State University, Stillwater, Oklahoma 74078-3072, USA}
\author{S. Chaturvedi}
\affiliation{School of Physics, University of Hyderabad, Hyderabad 500046, India}


\begin{abstract}
We present a self contained formalism  modelled after the Brownian motion of a quantum harmonic oscillator 
for describing the performance of microscopic Brownian heat engines like Carnot, Stirling and Otto engines. 
Our theory, besides reproducing the standard thermodynamics results in the steady state enables permits 
us to study the role  dissipation plays in determining the efficiency of Brownian heat engines under 
actual laboratory conditions. In particular, we analyse in detail the dynamics associated with 
decoupling  a system in equilibrium with one bath and recoupling it to another bath and obtain exact 
analytical results which are shown to have significant ramifications on the efficiencies of engines involving such a 
step. We also develop a simple yet powerful technique for computing corrections to the steady state results 
arising from finite operation time  and use it to arrive at the thermodynamic complementarity relations for 
various operating conditions and also to compute the efficiencies of the three engines cited above at maximum power.
Some of the methods and techniques and exactly solvable models presented here are interesting in their own right and, in our opinion, would find useful applications in other contexts as well.
\end{abstract} 
\maketitle

\section{Introduction}
In recent years there has been an upsurge of interest in the interface 
between thermodynamics and quantum mechanics \cite{1}, macro and nano/micro \cite{2}, 
leading to a thorough reexamination of the basic concepts and principles of 
thermodynamics with ramifications in biological processes and soft condensed 
matter systems. New paradigms for notions of work, thermal machines etc. have emerged 
that provide deep insights into thermodynamics which in turn help enlarge its scope 
far beyond that envisaged originally and open up new possibilities \cite{3}. 
These developments force one to formulate questions concerning efficiencies of various 
heat engines using appropriate microscopic considerations \cite{4}. Indeed, one has even 
started considering quantum heat engines which in principle have efficiency larger than 
Carnot efficiency though such cases require non equilibrium steady states. 
Such steady states can be reached via the use of the coherent laser fields or via quantum interference 
effects \cite{5}-\cite{6}. In a very recent experiment \cite{7} realized a microscopic 
Brownian heat engine \cite{8}-\cite{10}. The most important ingredient both in the work of 
Blickle and Bechinger \cite{7} as well as  in the proposals of Scully and collaborators
\cite{5} is the possibility that all the relevant parametrs can be very well controlled 
experimentally and thus the heat engine cycle can be precisely realized and it becomes 
desirable to have exactly soluble models of microscopic heat engines.
 
In view of the way the experiments are carried out we need a fully dynamical model which should account 
for the way the system parameters like potentials or the external parameters like temperature are changed. 
Further the behavior of the engine should depend on various time scales foe example the time taken to 
reach equilibrium state. Such a time would depend on the scales of the damping in the system. Motivated by these requirements we develop in the present work an exactly soluble model of a microscopic Brownian heat engine. 
The model that we present is fully quantum mechanical. Our model enables us to examine 
many different possible experimental scenarios- (a) low temperature behavior where 
quantum effects are likely to dominate, (b) behavior under different relaxation 
conditions-for example the system could be underdamped or overdamped, (c) possibilities 
for the system 
to pass through nonequilibrium stages depending on the rate of change of the external 
parameters, (d) nonequilibrium conditions because the experimental time scales are 
smaller than the time it takes for the system to reach steady state. Our formulation is 
based on the Wigner function and quantum Langevin equations for an harmonic oscillator 
whose frequency is modulated  in time. We also assume that 
the temperature of the environment is  time dependent as well. These time dependences are 
needed to implement the heat engine cycle realistically. We calculate the time dependent
 Wigner function, all the fluctuation parameters and the entropy. These enable us to 
calculate thermodynamic quantities like work, heat and internal energy.

A brief outline of the work is as follows. In Section II  for later reference we briefly 
 recapitualate the expressions for the efficiencies, both classical as well as quantum, 
for the three engines based on standard thermodynamic considerations with usual assumptions regarding 
the speed with which the various steps are carried out. In Section III we present a qua
ntum thermodynamic frame work based on Wigner phase space description for 
quantum systems which contains the classical framework as a limiting case. In Section IV
we develop a general set up for computing various quantitities of interest and give two models 
of frequency modulation where the relevant equations are amenable to exact analytical 
results. In Section V we consider the situation when the diffusion constant is varied linearly 
and in Section VI analyse its ramifications on the efficiencies of Brownian heat engines. In Section VII  we develop a systematic scheme for computing finite time 
corrections to the efficiencies of classical and quantum Brownian motors and then use these results 
in Section VII to examine the role they play in determining the
efficiency  of the Stirling engine at maximum power. Section IX contains our concluding remarks and further 
outlook.

As noted above our working model for a heat engine is based on a quantum harmonic 
oscillator with a frequency $\omega$ interacting with a thermal bath at temperature $T$. 
Three typical engines that have been discussed extensively in the literature based on 
varying $\omega$ and $T$ appropriately are (a) The Stirling (b) Carnot and (c) Otto engines.
Their cycles adapted to the harmonic oscillator model are schematically given below:
\vskip3mm
\noindent
{\bf Stirling Engine}~\cite{7}
\begin{align}
\label{1a}
&\begin{array}{cccccc}
\omega_1,T_c~&  &\text{Isothermal}& &&\omega_2,T_c\\
& 3 &\longrightarrow& 4&&\\
&  &\tau_c& &&\\
 \text{Isochoric}&\uparrow&&\downarrow&&\text{Isochoric}\\
&  &\tau_h& &&\\
& 2&\longleftarrow&1 &&\\
\omega_1,T_h~& &\text{Isothermal}& &&\omega_2,T_h\\
&&&&&\\ \end{array} \\
&~~~~~~~~~~~~~~~\omega_2>\omega_1,~T_h>T_c~~~~~~~~~~,
\nonumber
\end{align}

\noindent
{\bf Carnot Engine}~\cite{9},\cite{11}
\begin{align}
\label{2a}
&\begin{array}{cccccc}
\omega_1,T_h~&  &\text{Isothermal}& &&\omega_2,T_h\\
& 1 &\longrightarrow& 2&&\\
&  &\tau_h& &&\\
 \text{Isentropic}&\uparrow&&\downarrow&&\text{Isentropic}\\
&  &\tau_c& &&\\
& 4&\longleftarrow&3 &&\\
\omega_4,T_c~& &\text{Isothermal}& &&\omega_3,T_c\\
&&&&&\\ \end{array}\\
&\omega_1>\omega_2>\omega_3>\omega_4,~T_h>T_c,~\beta_h\omega_2=\beta_c\omega_3,
\beta_h\omega_1=\beta_c\omega_4~~~~~~~~~~~~~~~~~,
\nonumber
\end{align}

{\bf Otto Engine}~\cite{12}
\begin{align}
\label{3a}
&\begin{array}{cccccc}
\omega_c,T_2~&  &\text{Isentropic}& &&\omega_h,T_h\\
& 4 &\longleftarrow& 3&&\\
&  &\tau_1& &&\\
 \text{Isochoric}&\downarrow&&\uparrow&&\text{Isochoric}\\
&  &\tau_2& &&\\
& 1&\longrightarrow&2 &&\\
\omega_c,T_c~& &\text{Isentropic}& &&\omega_h,T_1\\
&&&&&\\ \end{array}\\
&\omega_h>\omega_c>,~T_h>T_c,~\beta_c\omega_c=\beta_1\omega_h,~\beta_h\omega_h=
\beta_2\omega_c,~~~.
\nonumber
\end{align}
Here, the $\tau$'s indicate the time taken to carry out the indicated step and $\beta$ stands for $1/K_BT$.
The calculations that we give in subsequent sections can be applied to any of these engines.

The  three prototype engines above thus involve  suitable combinations of the following 
three steps:~(a) isothermal i.e. $\omega$ changing, $T$ fixed or (b) isochoric 
i.e. $\omega$ held fixed, $T$ changing or (c) isentropic 
i.e both $\omega$ and $T$ changing with $\omega/T$ fixed and one needs an appropriate 
formalism to compute the efficiencies under specific physical circumstances in which 
these steps are actually executed in an experiment. The present work has this as its 
major objective. Our principal results include (i) development of a  self contained formalism 
for computing efficiencies of Brownian engines both in the classical as well as quantum 
conexts (ii) an exact analysis of the role of damping in the process of coupling the system to a bath at a higher temperature and its influence on the performance of the Stirling engine (iii)  computation of the 
irreversible heat in isothermal processes and the derivation of complementarity relations 
(iv) a detailed analysis of the role of damping as well as finite time corrections on the 
efficiency of the Stirling engine at maximum power.

\section{Steady State Efficiencies from Thermodynamics}
To set the notation and for later reference we assemble here the standard thermodynamic considerations that enable 
us to compute the efficiencies  for the three engines listed above both in classical and as well as 
in the quantum contexts. These are: 
\begin{enumerate}
\item  the thermodynamic conservation law $\Delta U=\Delta Q-\Delta W;$ where 
$ \Delta U$:
Change in the internal energy $U$;~~$\Delta Q$:~Heat absorbed by the system;
$\Delta W$:~Work done by the system
\item  $\Delta Q$ in an isentropic process 
~$ a \rightarrow b =0$,
\item work done in an isochoric process~$a \rightarrow b =0$,
\item work done in an isothermal process~ $a \rightarrow b = -[F(b)-F(a)]$ where $F$ 
denotes the free energy of the system,
\item the expressions for $U$ and $F$ for the harmonic oscillator:
\begin{align}
U=\begin{array}{l} 1/\beta,~ ~~\beta\equiv\frac{1}{K_BT}~~~~~~~~~~~~~~~~~~~~~~~~~~~~~~~~
(\text{Classical})\\
\\\hbar\omega[n(\omega,T)+1/2],~ n(\omega, T)\equiv\dfrac{1}{(e^{\beta\hbar\omega}-1)}~~~
(\text{Quantum})\end{array},
\end{align}
\begin{align}
F(\omega,T)=\begin{array}{l}\dfrac{1}{\beta}\ln(\beta\hbar\omega)~~~~~~~~~~~~~~~~~~~~~~~~~~~~~(\text{Classical})\\
\\ \dfrac{1}{\beta}\ln(2\sinh(\beta\hbar\omega/2))
~~~~~~~~~~~~~~~~(\text{Quantum})\end{array},
\end{align} 
\item the expression for the entropy of a classical harmonic oscillator 
\begin{equation}
 S=K_B\left[1+\ln\left(\dfrac{1}{\beta\hbar\omega}\right)\right].
\end{equation}
In an isothermal process one has
\begin{equation}
 \Delta Q=\Delta W=T\Delta S.
\end{equation}

In the quantum case, for the thermal states $\rho_{th}$, 
\begin{equation}
 \rho_{th}=\dfrac{e^{-\beta\hat{H}}}{\text{Tr}[e^{-\beta\hat{H}}]},
\end{equation}
where $\hat{H}$ denotes the hamiltonian for a quantum harmonic oscillator, one has for the von Neumann entropy
\begin{equation}
 S=K_B[(n(\omega,T)+1)\ln(n(\omega,T)+1)-n(\omega,T)\ln n(\omega,T)].
\end{equation}
\end{enumerate}
With this preparation we now proceed to compute the efficiencies of the three engines mentioned
earlier both in the classical as well as quantum cases. These would then be compared with the 
results obtained from the microscopic theory developed later.
\subsection{Stirling Engine} 
The efficiency $\eta_s$ of the Stirling engine is defined as
\begin{equation}
\eta_s=\dfrac{\text{Work done by the system}}{\text{Heat flow into the system at}~T_h}.
\nonumber
\end{equation}
\vskip2mm
\noindent
{\bf Classical}

In the classical case the work done by the engine is given by
\begin{align}
\Delta W_{1\rightarrow 2}&+\Delta W_{3\rightarrow 4}
=-[F(\omega_1,T_h)-F(\omega_2,T_h)]-[F(\omega_2,T_c)-F(\omega_1,T_c)]\nonumber\\
&=K_B(T_h-T_c)\ln\left(\dfrac{\omega_2}{\omega_1}\right),
\end{align}
and the heat absorbed at $T_h$ by
\begin{align}
\Delta W_{1\rightarrow 2}&+{\Delta U}_{1\rightarrow 2}+{\Delta U}_{4\rightarrow 1}
=-[F(\omega_1,T_h)-F(\omega_2,T_h)]+0+\dfrac{1}{2}(\dfrac{1}{\beta_h}-\dfrac{1}{\beta_c})\nonumber\\
&=K_BT_h\ln\left(\dfrac{\omega_2}{\omega_1}\right)+\dfrac{1}{2} K_B(T_h-T_c).
\label{11a}
\end{align}
[ Note the factor of half in the second term on the RHS of the above equation.
We will return to  this later.]
Hence
\begin{equation}
\eta_s^{\text{cl}}=\dfrac{\eta_c}{1+\eta_c/(\ln(\dfrac{\omega_2^2}{\omega_1^2})},~~
\eta_c=1-\dfrac{T_c}{T_h}.
\end{equation}
\noindent
{\bf Quantum}

Proceeding as before and using the expressions for $U$ and $F$ appropriate to the quantum case, we have
for the work done
\begin{align}
\Delta W_{1\rightarrow 2}&+\Delta W_{3\rightarrow 4}
=-[F(\omega_1,T_h)-F(\omega_2,T_h)]-[F(\omega_2,T_c)-F(\omega_1,T_c)],\nonumber\\
&=K_BT_h \ln\left(\dfrac{\sinh(\beta_h\hbar\omega_2/2)}{\sinh(\beta_h\hbar\omega_1/2)}\right)     
 -K_BT_c\ln\left(\dfrac{\sinh(\beta_c\hbar\omega_2/2)}{\sinh(\beta_c\hbar\omega_1/2)}\right),
\end{align}
and  for the heat absorbed at $T_h$
\begin{align}
 \Delta W_{1\rightarrow 2}&+{\Delta U}_{1\rightarrow 2}+{\Delta U}_{4\rightarrow 1}
=-[F(\omega_1,T_h)-F(\omega_2,T_h)]\nonumber\\&+[\hbar\omega_1(n(\omega_1,T_h)+1/2)
-\hbar\omega_2(n(\omega_2,T_h)+1/2)]\nonumber\\&+\dfrac{1}{2}\left([\hbar\omega_2(n(\omega_2,T_h)+1/2)]
-[\hbar\omega_2(n(\omega_2,T_c)+1/2)]\right),
\end{align}
and hence
\begin{align}
&~~~~~~~~~~~~~~~~~~~~~~~~~~~~~~~~~\eta_s^{\text{q}}=\dfrac{1-Y/X}{1+Z/X}~~~,\nonumber\\
&~~~~~~X=\ln\left(\dfrac{\sinh(\beta_h\hbar\omega_2/2)}{\sinh(\beta_h\hbar\omega_1/2)}
\right),~~~Y= \dfrac{\beta_h}{\beta_c}\ln\left(\dfrac{\sinh(\beta_c\hbar\omega_2/2)}
{\sinh(\beta_c\hbar\omega_1/2)}\right)~~,\\
&Z=\dfrac{
\beta_h}{2}\left[\hbar\omega_1\coth\left(\beta_h\hbar\omega_1/2\right)-\dfrac{\hbar\omega_2}{2}\{\coth\left(\beta_h\hbar\omega_2/2\right)+
\coth\left(\beta_c\hbar\omega_2/2\right)\}\right].\nonumber
\label{11}
\end{align}

In the limit $\beta\omega~<<~1$ $\eta_s^{\text{q}}$ goes over to the classical efficiency $\eta_s^{\text{cl}}$ 
as expected.

\subsection{Carnot Engine}
For the Carnot engine the efficiency defined as before 
\begin{equation}
\eta_c=\dfrac{\text{Work done by the System}}{\text{Heat flow into the system at}~T_h},
\nonumber
\end{equation} 
turns out to be the same in both classical and quantum cases and is given by 
\begin{equation}
 \eta_c^{\text{cl}}=\eta_c^{\text{q}}= \eta_c=\left(1-\dfrac{T_c}{T_h}\right).
\end{equation}
\subsection{Otto Engine}
Here again the efficiency defined as
\begin{equation}
\eta_o=\dfrac{\text{Work done by the System}}{\text{Heat flow into the
 system during}~2\rightarrow 3},
\nonumber
\end{equation} 
turns out to be the same in both quantum and classical cases and is given by
 \begin{align}
 \eta_o^{\text{cl}}=\eta_o^{\text{q}}& =1-\dfrac{U(4)-U(1)}{U(3)-U(2)}\nonumber\\
&=\left(1-\dfrac{\omega_c}{\omega_h}\right).
\end{align} 
The expressions for efficiencies for the the three engines, realized here through a harmonic oscillator 
by appropriate changes of its frequency ( or equivalently its `spring constant') and the ambient temperature,  
hold for idealized operating conditions  as stipulated in equilibrium thermodynamics. These, for instance, 
demand that the isothermal changes of frquency involved in the Stiring or the Carnot cycles  
be carried out quasistatically i.e. so slowly that at each instance the oscillator remains in the state of 
equilibrium at that temperature and frequency. Such conditions are hardly ever met in practice and particularly 
in the light of the experimental work reported in \cite{7} there is an obvious need for developing a framework 
which brings into play aspects of approach to equilibrium , both in classical and quantum contexts, and 
is capable of furnishing a self-contained scheme for computing the efficiencies under realistic conditions. We 
develop such a scheme in the next section.

\section{Efficiencies beyond the steady state: A dynamical model}
To go beyond the standard  thermodynamic assumptions regarding the rate at which 
which various steps in a heat engine are carried out so that one can evaluate the 
performance of an engine under actual  laboratory conditions we need a framework which treats the system modelling the 
engine as an open system and permitting proper inclusion of  dissipative effects and the possibility of varying 
the ststem potential and the ambient temperature. In the present context, such a framework is provided by 
the dynamics of a quantum Brownian oscillator of frequency $\omega$ in contact with a 
heat bath at temperature $T$.is described by the master equation \cite{13}
\begin{align}
 \dfrac{\partial}{\partial t}\rho&=-\dfrac{i}{\hbar}[\hat{p}^2/2m+\dfrac{1}{2}m\omega^2\hat{q}^2, \rho]-
 \dfrac{2\kappa m\omega}{\hbar}(n(\omega,T)+1/2)([\hat{q},[\hat{q},\rho]])\nonumber\\&
~~~~~~~~~~~~~- \dfrac{i\kappa}{\hbar}([\hat{q},\{\hat{p},\rho\}]),
\end{align}
where $\hat{q}$ and $\hat{p}$ are denote the position and 
momentum operators obeying the commutation relations $[\hat{q},\hat{p}]=i\hbar$.

 For reasons given later it 
proves expedient to transcribe the quantum dynamics described by the master equation using 
the Wigner phase space description of quantum systems \cite{14},\cite{15}  which 
associates with a density operator $\rho$ a phase space function $W(q,p)$ of classical 
variables $q,p$ as follows:
\begin{eqnarray}
\widehat{\rho}\mapsto W_{\widehat{\rho}}(q,p) &=& {\rm Tr}\left\{ \widehat{\rho}~
\widehat{W}(q,p)\right\}~;\nonumber\\
\widehat{W}(q,p) &=&
\dfrac{1}{(2\pi\hbar)}\int\limits_{-\infty}^{\infty}dq^{\prime}~|q+\dfrac{1}{2}q^{\prime}
\rangle
\langle
q-\dfrac{1}{2}q^{\prime}|~e^{i\;pq^{\prime}/\hbar},
\end{eqnarray}

We note here that we use the Wigner phase space description in preference to other phase 
space descriptions for two reasons: (a) it is the only 
one that that maps the quantum mechanical average of a product of two operators to the 
phase space average of the corresponding Wigner functions (b) its moments $<q^mp^n>$ 
correspond to quantum averages of of the symmetrised operator $(\hat{q}^m\hat{p}^n)_S$. 
For example $<q^2p>$ corresponds to the expectation value of the operator 
$(\hat{q}^2\hat{p}+\hat{q}\hat{p}\hat{q}+\hat{p}\hat{q}^2)/3$.

Use of the Wigner description turns the master equation into a Fokker-Planck equation for 
$W(q,p)$ \cite{15}
\begin{equation}
 \dfrac{\partial}{\partial t}W(q,p,t)=\left[-\dfrac{\partial}{\partial q}
\left(\dfrac{p}{m}\right)
+
\frac{\partial}{\partial p}\left(2\kappa p+
\left(\frac{\partial V(q,a)}{\partial q}\right)\right)
+D\dfrac{\partial^2}{\partial p^2}\right]W(q,p,t), \nonumber\\
\end{equation}
where
\begin{equation}
 V(q,a)=\dfrac{1}{2}aq^2,~~~~a\equiv m\omega^2,
\label{2}
\end{equation}
and 
\begin{equation}
 D=2m\hbar\omega\kappa(n(\omega,T)+\dfrac{1}{2}),~~~n(\omega,T)=(e^{\beta\hbar\omega}-1)^{-1}.
\label{3}
\end{equation}
In the following the parameter $a$, the `spring constant', will be taken to be controlled externally.

The Langevin equations equivalent to the above FPE read:
\begin{align}
 &\dot{q}=\dfrac{p}{m} ,\label{4}\\
&\dot{p}=-2\kappa p-\dfrac{\partial}{\partial q}V(q,a)+f(t),\label{5}\\
&~~~~~<f(t)f(t^\prime)>=2D\delta(t-t^\prime).
\label{6}
\end{align}
The Langevin equations $(\ref{4})-(\ref{6})$~ lend themselves to a nice thermodynamics 
intepretation \cite{8}:
Rewriting $(\ref{5})$ as
\begin{equation}
 -(-2\kappa p+f(t))+\dot{p}+\dfrac{\partial}{\partial q}V(q,a)=0,
\end{equation}
and multiplying it by $dq$ and using
\begin{equation}
 dV=\frac{\partial V(q,a)}{\partial q}dq+\frac{\partial V(q,a)}{\partial a}da,
\end{equation}
one obtains
\begin{equation}
 -(-2\kappa p+f(t))dq +d(p^2/2m+V(q,a))-\frac{\partial V(q,a)}{\partial a}da.
\label{20}
\end{equation}
The three terms in the above equation may now be identified in an intuitively plausible 
manner as:
\begin{equation}
 d{\cal Q}=(-2\kappa p+f(t))dq,~~d{\cal U}=d(p^2/2m+V),~~~d{\cal W}=-\frac{\partial V(q,a)}{\partial a}da,
\label{21}
\end{equation}
leading to the energy balance equation:
\begin{equation}
 -d{\cal Q}+d{\cal U}+d{\cal W}=0,
\end{equation}
with $d{\cal Q}$ ( -$d{\cal Q}$) understood as the heat flow into of (out) the system and 
$d{\cal W}$
(-$d{\cal W}$) as the work done by (on) the system. The stochastic 
averages of these quantities denote by 
$dQ,dU ~\text{and} ~dW$ respectively  relate directly to the corresponding 
thermodynamic quantities and capture the thermodynamic conservation laws. 
This self-contained approach is clearly more microscopic than thermodynamics as it 
provides a framework for computing not only the averages of these quantities but 
their probability distributions as well. 

We note here that while it is certainly possible to transcribe the master equation 
dynamics directly into equivalent quantum Langevin equations for the operators $\hat{q}$ and $\hat{p}$ 
but owing to their noncommutativity the crucial step $(\ref{20})$ needed to obtain a clear 
thermodynamic interpretation of the such Langevin equations would now involve terms like $\hat{p}.d\hat{q}$ 
and would therefore be fraught with ordering ambiguities. 

The scheme described above for computing $dQ, ~dU$ and $dW$ together with the 
expression for von Neumann entropy 
\begin{equation}
 S=K_B[(\sigma+1)\ln(\sigma+1)-\sigma\ln\sigma],~~\sigma=\sqrt{\text{Det}[{\cal V}]}-\frac{1}{2} ,
\end{equation}
for Gaussian states \cite{16} ( which is what we would 
exclusively deal with) i.e. states $\rho$ for which the Wigner distribution  is a Gaussian: 
\begin{equation}
 W(q,p)=\frac{1}{\sqrt{(2\pi)^2\text{Det}[{\cal V}]}}\exp\left[-\frac{x^T{\cal V}x}{2\text{Det}[{\cal V}]}\right],~~~x\equiv\left(\begin{array}{c}
q\\p\end{array}\right),
\end{equation}
provide all that we need for the considerations below. Here $V$ stands for the variance matrix,
\begin{equation}
 {\cal V}=\left(\begin{array}{cc}
<q^2>&<qp>\\<qp>& <p^2>\end{array}\right),
\end{equation}
and $<\cdot>$ denote averages with respect to the Wigner distribution. The uncertainty relations 
require that $\sigma$ be positive. Note that the set of Gaussian states contains the set of  harmonic oscillator 
thermal states $\rho_{th}$ as a special case.

Before proceeding further it is instructive to check that the Wigner description above 
together with the thermodynamic interpretation implied by $(\ref{21})$ in the steady state 
does indeed reproduce the results given earlier for the efficiencies of the three engines 
using standard thermodynamic considerations. Thus, for instance, calculation  of 
the efficiency of the Stirling engine involves computing ${\Delta W}_{1\rightarrow 2}$
${\Delta  U}_{1\rightarrow 2}$, ${\Delta U}_{4\rightarrow 1}$ which in the present 
framework are given by \begin{align}
&\Delta{ W}_{1\rightarrow 2}=\int_{1}^{2} d{\cal W}
= -\int_{\omega_2}^{\omega_1} m\omega <q^2>_{T=T_c} d\omega,\\
&\Delta{ W}_{3\rightarrow 4}=\int_{3}^{4} d{\cal W}
= -\int_{\omega_{1}}^{\omega_{2}} m\omega <q^2>_{T=T_h} d\omega,\\
&\Delta{U}_{1\rightarrow 2}=\int_{1}^{2} d{ U }
=\left(\frac{<p^2>}{2m}+\frac{1}{2}m\omega^2<q^2>\right)_2-
\left(\frac{<p^2>}{2m}+\frac{1}{2}m\omega^2<q^2>\right)_1,\\
&\Delta{ U}_{4\rightarrow 1}=\int_{4}^{1} d{ U }
=\left(\frac{<p^2>_1}{2m}-
\frac{<p^2>_4}{2m}\right).
\label{37a}
\end{align}

Further, from the FPE or the Langevin eqns it follows that in the steady state 
\begin{equation} 
 <p^2>=D/2\kappa,~~~~~~~~m\omega^2<q^2>=D/2m\kappa.
\end{equation}
These on using $D=2m\kappa\omega(n(\omega,T)+1/2)$ then give
\begin{align}
 &\Delta{ W}_{1\rightarrow 2}=
-\int_{\omega_{2}}^{\omega_{1}} [n(\omega,T_h+1/2] d\omega\nonumber\\
&~~=K_BT_h\ln\left(\frac{\sinh(\beta_h\omega_2/2)}
{\sinh(\beta_h\omega_1/2)}\right)=F(1)-F(2),\\
&\Delta{ W}_{3\rightarrow 4}
= -\int_{\omega_{1}}^{\omega_{2}} [n(\omega,T_c)+1/2]  d\omega\nonumber\\
&~~=-K_BT_c\ln\left(\frac{\sinh(\beta_c\omega_2/2)}
{\sinh(\beta_c\omega_1/2)}\right)=[F(3)-F(4)],\\
&\Delta{ U}_{1\rightarrow 2}=
[\omega_{1}(n(\omega_{1},T_h)+\frac{1}{2})
-[\omega_2(n(\omega_{2},T_h)+\frac{1}{2})],\\
&\Delta{U}_{4\rightarrow 1}=\frac{1}{2}\left(
[\omega_{2}(n(\omega_{2},T_h)+\frac{1}{2})]-[\omega_2(n(\omega_{2},T_c)
+\frac{1}{2})]\right),
\end{align}
which are the same expressions as before and therefore one recovers the expression for 
efficiency given in Section II. ( Note here that in computing 
$\Delta{ U}_{4\rightarrow 1}$ we considered only the contribution from $<p^2>$ and 
not from $<q^2>$, a question that will be examined in greater detail later.)

\section{Quantum dynamics under time dependent changes of temperature and potential}

We have seen in the previous  section that the Langevin equations equivalent to the 
Fokker-Planck equation obeyed by the Wigner distribution lend themselves to a direct 
and transparent thermodynamic interpretation and that this interpretation in 
the steady state limit reproduces the standard thermodynamic results. To prepare ground 
for going beyond the steady state limit we now analyse the structure of the solutions 
of the Langevin equations at hand allowing for arbitrary time dependence in the 
potential and the diffusion coefficients and apply this framework to arrive at the 
exact solutions of the Langevin equations for three fairly realistic models.

The Langevin equations which in the present case are linear stochastic 
equations with additive noise may be solved to yield :
\begin{equation}
\left( \begin{array}{c}q(t)\\p(t)\end{array}\right)=M(t) \left(\begin{array}{c}
q(0)\\p(0)\end{array}\right)+\int_0^tdt^\prime M(t)M(t^\prime)^{-1}
\left(\begin{array}{c}0\\\sqrt{2D(t^\prime)}f(t^\prime)\end{array}\right),
\end{equation}
where
\begin{equation}
 M(t)\equiv \left(\begin{array}{cc}u(t)& v(t)\\ m\dot{u}(t)& m\dot{v}(t)\end{array}\right),
\end{equation}
solves the homogeneous equations 
\begin{equation}
\frac{d}{dt}\left(\begin{array}{c}q(t)\\p(t)\end{array}\right)=
\left(\begin{array}{cc}1/m& 0\\ -m\omega^2&-2\kappa\end{array} \right)
 \left(\begin{array}{c}q(t)\\p(t)\end{array}\right).
\end{equation}
From $(\ref{4})$ for the variance matrix 
\begin{equation}
{\cal V}(t)\equiv \left(\begin{array}{cc}<q^2(t)>& <q(t)(p(t)>\\ <q(t)p(t)>& <p^2(t)>\end{array}\right),
\label{19}
\end{equation}
one has 
\begin{equation}
{\cal V}(t)=M(t)[~{\cal V}(0)+\int_{0}^{t}dt^\prime M^{-1}(t^\prime)
\left(\begin{array}{cc}0& 0\\ 0& 2D(t^\prime)\end{array}\right)M^{T-1}(t^\prime)~]M^{T}(t).
\end{equation}
It is therefore clear that finding explicit solutions for the variances in situations 
where both $\omega$ and $T$ depend on time depends on our ability to solve for $M(t)$. We 
list below three physically meaningful cases where this is indeed possible.
\vskip3mm
\noindent
 {\bf Case I: $\omega $ independent of time }
\vskip3mm
For this familiar case the functions $U(t)$ and $V(t)$ 
which determine the matrix $M(t)$ are explicitly given by 
\begin{equation}
 u(t)=\frac{(\lambda_+e^{-\lambda_-t}-\lambda_-e^{-\lambda_+t})}{(\lambda_+-\lambda_-)},~
v(t)=\frac{(e^{-\lambda_-t}-e^{-\lambda_+t})}{m(\lambda_+-\lambda_-)};~\lambda_{\pm}=
\kappa\pm\sqrt{\kappa^2-\omega^2}.
\label{47}
\end{equation}
Further, owing to time translation available in this case, we have 
 $M^{-1}(t)=M(-t),~~M(t)M(t^\prime)=M(t+t^\prime)$, and $(\ref{19})$
simplifies to 
\begin{equation}
{\cal V}(t)=M(t){\cal V}(0)M^{T}(t)+\int_{0}^{t}dt^\prime M(t^\prime)
\left(\begin{array}{cc}0& 0\\ 0& 2D(t-t^\prime)\end{array}\right)M^{T}(t^\prime).~
\label{48}
\end{equation} 
\vskip3mm
\noindent
{\bf Case II : $\omega^2(t)=\omega_0^2\left(1+\frac{\mu t}{T}\right),~0\leq t\leq T $ }
\vskip3mm
In this case investigated in \cite{17}, the functions $u(t)$ and $v(t)$ 
in the range $0\leq t\leq T $
 are given by 
\begin{align}
& u(t)=\left[\frac{{f_+}(t)\dot{f}_-(0)-f_-(t)\dot{f}_+(0)}
{f_+(0)\dot{f}_-(0)-f_-(0)\dot{f}_+(0)}\right] 
~, v(t)=m\left[ \frac{f_+(t)f_-(0)-f_-(t)f_+(0)}
{\dot{f_+}(0)f_-(0)-\dot{f_-}(0)f_+(0)}\right],
\\
&f_{\pm}(t)=e^{-\kappa t}(t+a)^{1/2}J_{\pm 1/3}
\left(\frac{2}{3}b^{1/2}(t+a)^{3/2}\right),~a=\left(1-\frac{\kappa^{2}}{\omega_0^2}
\right)\frac{T}{\mu},~b=\frac{\omega_0^2\mu}{T}.
\label{23}
\end{align}

\vskip3mm
\noindent
{\bf Case III : $\omega^2(t)=\omega_0^2 e^{\frac{\mu t}{T}}, ~0\leq t\leq T $}
\vskip3mm
In this case the functions $f_+(t)$ and $f_-(t)$ in $(\ref{23})$ are again given in terms of 
Bessel functions as 
\begin{equation}
 f_{\pm}(t)=e^{-\kappa t}J_{\pm \alpha}
\left(ae^{\frac{\mu t}{2T}}\right),~a=\frac{2T\omega_0}{\mu}~ \alpha
=\frac{2T\kappa}{\mu}.
\end{equation}

Having dealt with some exactly solvable cases where the frequency is changed in a specific 
way but the temperature may be varied arbitrarily, we now illustrate how the formalism 
developed above lends itself to useful exact or approximate calculations leading to 
finite time corrections.
\section {Linear variation of the diffusion constant}
 We first consider the case in which the the harmonic oscillator with frequency 
$\omega$ is in equilibrium with a bath at temperature $T_0$ characterized by a diffusion 
constant $D_0$. With $\omega$ held fixed, the the diffusion 
coefficient is changed linearly from its initial value $D_0$ appropriate to 
temperature $T_0$ to its final value $D_1$ appropriate to temperature $T_1$
in a time $\tau$ and then kept at that value thereafter. 
\begin{align}
 &\omega :\text{held fixed}\nonumber\\
&D(t)=\begin{array}{l}D_0+(D_1-D_0)\dfrac{t}{\tau},~0\leq~t~\leq~\tau\\
\\ D_1,~~~~t>\tau.\end{array} 
\end{align}
This situation pertains to the isochoric step in the Brownian engines and is relevant for 
discussions on aspects of decoupling the system from a heat bath at one temperature and 
recoupling it to another heat bath at a different temperature. 
 
For the case at hand, with ${\cal V}(0)$ chosen to be the variance matrix corresponding 
to the oscillator being at equilibrium with the bath at temperature appropriate to 
$D_0$ 
\begin{equation} 
 {\cal V}(0)=\left(\begin{array}{cc}\dfrac{D_0}{2\kappa m}&0\\0&\dfrac{D_0}{2\kappa m^2 \omega^2}
\end{array}\right),
\end{equation}
we have from $(\ref{48})$
\begin{align}
 &<q^2(t)> =\frac{D_0}{2\kappa}\left(\frac{u^2(t)}{m^2\omega^2}+v^2(t)\right)+2
\int_0^tdt^\prime v^2(t-t^\prime)D(t^\prime),\\
&<q(t)p(t)> =2\int_0^tdt^\prime v(t-t^\prime) \dot{v}(t-t^\prime)D(t^\prime),\\
&<p^2(t)> =\frac{mD_0}{2\kappa}\left(\frac{u(t)\dot{u}(t)}{m^2\omega^2}+v(t)\dot{v}(t)
\right)+2m^2
\int_0^tdt^\prime \dot{v}^2(t-t^\prime)D(t^\prime).
\end{align}
Using the relations
\begin{align}
& v^2=-\frac{1}{2\kappa}\frac{1}{2}\frac{d}{dt}\left(\frac{u^2}{m^2\omega^2}+v^2\right),\\
& v\dot{v}=-\frac{1}{2\kappa}\frac{d}{dt}\left(\frac{u\dot{u}}{m^2\omega^2}+v\dot{v}\right),
\\
& \dot{v}^2=-\frac{1}{2\kappa}\frac{1}{2}\frac{d}{dt}\left(\frac{\dot{u}^2}{m^2\omega^2}+
\dot{v}^2\right),
\end{align}
which follow from 
\begin{align}
 \dot{v}=-2\kappa v+\frac{u}{m};~~\dot{u}=- m\omega^2 v,
\end{align}
we obtain for $t>\tau$
\begin{align}
& <q^2(t)>=\alpha(t) <q^2>_0 +(1-\alpha(t)) <q^2>_1\label{61}\\
&<p^2(t)>= \beta(t) <p^2>_0 + (1-\beta(t)) <p^2>_1.\label{62}
\end{align}
where
\begin{align}
&\alpha(t)=\frac{1}{\tau}\int_{t-\tau}^{t}dt^\prime[m^2\omega^2v^2(t^\prime)+u^2(t^\prime)],
\label{63}\\
&\beta((t)=\frac{1}{\tau}\int_{t-\tau}^{t}dt^\prime[m^2\dot{v}^2(t^\prime)+\frac{\dot{u}^2
(t^\prime)}{\omega^2}].\label{64}
\end{align}
In the limit $t\rightarrow \infty$ limit both $\alpha(t)$ and $\beta(t)$ go to zero and hence 
$<q^2>$ and $<p^2>$ assume their respective equilibrium values. The parameters $\alpha(t)$ and $\beta(t)$ thus interpolate between the initial and the final equilibrium values of $<q^2>$ and $<p^2>$ and quantify the approach
to equilibrium. In the following we consider the case when $t=\tau$ i.e. the situation that obtains 
immediately after the bath has reached the state characterized by the final value of the diffusion coefficient. 
Evidently as far as the system is concerned we are  dealing here with a non equilibrium  state as the system 
has not yet had time to equilibriate with the `final' bath. 

Putting $t=\tau$ in $(\ref{63})$ and $(\ref{64})$ and denoting $\alpha(\tau)$ and $\beta(\tau)$ simply as $\alpha$ and $\beta$ we obtain on substituting for $u$ and $v$ from $(\ref{47})$ and carrying out the relevant integrals
\begin{align}
&\alpha=\frac{1}{(x-y)^2}\left[(x+y)\left(\frac{x}{2y}(1-e^{-2y})+\frac{y}{2x}(1-e^{-2x})\right)
-\frac{4xy}{(x+y)}(1-e^{-(x+y)})\right],\\
&\beta=\frac{1}{(x-y)^2}\left[(x+y)\left((1-\frac{(e^{-2y}+e^{-2x})}{2}\right)
-\frac{4xy}{(x+y)}(1-e^{-(x+y)}\right],
\end{align}
where $x=[\kappa+\sqrt{\kappa^2-\omega^2}]\tau, y=[\kappa-\sqrt{\kappa^2-\omega^2}]\tau$. 
We now examine the behaviour of $\alpha$ and $\beta$ in the overdamped and weak dissipation 
regimes respectively.
\vskip5mm
\noindent 
{\bf Overdamped Case}
\vskip3mm
\noindent 
In the overdamped regime i.e.$\kappa>>\omega, x\approx 2\kappa\tau>>1, 
y\approx\frac{\omega^2\tau}{2\kappa}<<1$ one finds that
\begin{align}
& \alpha\approx\frac{1-e^{-2y}}{2y}\rightarrow 1~\text{as}~y\rightarrow 0,\\
&\beta\approx\frac{1-e^{-2x}}{2x}\rightarrow 0~\text{for}~x>>1.
\end{align}
\vskip5mm
\noindent 
{\bf Underdamped Case}
\vskip3mm
\noindent 
On the other hand, in the weak dissipation regime  $\kappa<< \omega, 
x\approx\kappa+i\omega , y\approx\kappa-i\omega$ and we have in the limit
$\kappa\tau\rightarrow 0$
\begin{align}
 \alpha&\approx\left[\frac{(1-e^{-2\kappa\tau})}{2\kappa\tau}+\kappa\tau\left(\frac{\sin\omega\tau}
{\omega\tau}\right)^2\right]\\
&\approx 1-\kappa\tau\left(1-\left(\frac{\sin\omega\tau}{\omega\tau}\right)^2\right),
\end{align}
\begin{align}
 \beta&\approx\left[\frac{(1-e^{-2\kappa\tau})}{2\kappa\tau}-\kappa\tau\left(\frac{\sin\omega\tau}
{\omega\tau}\right)^2\right]\\
&\approx 1-\kappa\tau\left(1+\left(\frac{\sin\omega\tau}{\omega\tau}\right)^2\right).
\end{align}
Note that $\alpha>\beta$ in both the cases. In fact this is always true -- it can easily 
be shown that with $x,y$ defined as before
\begin{equation}
 \alpha-\beta=\frac{x+y}{x-y}\int_0^1dt \left[e^{-2xt}-e^{-2yt}\right],
\end{equation}
and hence $\alpha>\beta$ by virtue of the fact that the integrand is always positive.
\section{Effect of time scales on Efficiencies of Brownian motors}
We recall that in the calculation of the efficiency of the Stirling engine from standard 
thermodynamic considerations presented in Section II, we had drawn attention to the factor 
of $1/2$ in the expression for $\Delta U_{4\rightarrow 1}$. Likewise in the computation of 
the Stirling engine using the quantum stochastic thermodynamics in the steady state we had 
noted that in computing  $\Delta U_{4\rightarrow 1}$ only  $<p^2>/2m$ contributes to 
$\Delta U_{4\rightarrow 1}$ and not $m\omega^2<q^2>/2$. This seemingly ad hoc prescription 
can now be understood at a deeper level in the light of the analysis in 
Section V leading to eqns. $(\ref{61})$-$(\ref{64})$. It is clear from the discussion 
therin that, in general, the expression for  $\Delta U_{4\rightarrow 1}$ should be taken 
to be   
\begin{align}
\Delta U_{4\rightarrow 1}&= (1-\beta)\left(\frac{<p^2>_1}{2m}-
\frac{<p^2>_4}{2m}\right)\nonumber\\&+(1-\alpha)m\omega^2\left(\frac{<q^2>_1}{2}- 
\frac{<q^2>_4}{2}\right),
\end{align}
where $\alpha$ and $\beta$ depend on various time scales involved. Indeed in the 
overdamped regime $\alpha\rightarrow 1$ and $\beta\rightarrow 0$ and one recovers the 
earlier results. The mystery behind the factor of $1/2$ in $(\ref{11a})$ and that behind 
retaining the contribution from $<p^2>/2m$ in $(\ref{37a})$ alone is thus resolved. 
On the other hand, in the weak dissipation regime 
where both $\alpha$ and $\beta$ are close to 1 the situation is very different and this has significant consequences 
for the  the relative magnitude of classical and quantum efficiencies under same operating conditions as 
discussed later. Further, since for a harmonic oscillator  $<p^2>/2m=m\omega^2<q^2>/2$ we can rewrite the above
equation as
\begin{align}
\Delta U_{4\rightarrow 1}= 2\mu\left(\frac{<p^2>_1}{2m}-
\frac{<p^2>_4}{2m}\right);~~~ \mu=1-\frac{\alpha+\beta}{2}.
\end{align}
Using this expression in the calculation of the classical and quantum efficiencies for the 
Stirlng engine given earlier respectively we obtain 
\begin{equation}
\eta_s^{\text{cl}}=\frac{\eta_c}{1+\eta_c\mu/\ln\left(\dfrac{\omega_2}{\omega_1}\right)}~~
,
\end{equation}
and
\begin{align}
&~~~~~~~~~~~~~~~~~~~~~~~~~~~~~~~~~\eta_s^{\text{q}}=\frac{1-Y/X}{1+Z/X}~~~,\nonumber\\
&~~~~~~X=\ln\left(\frac{\sinh(\beta_h\hbar\omega_2/2)}{\sinh(\beta_h\hbar\omega_1/2)}
\right),~~~Y= \frac{\beta_h}{\beta_c}\ln\left(\frac{\sinh(\beta_c\hbar\omega_2/2)}
{\sinh(\beta_c\hbar\omega_1/2)}\right)~~,\\
&Z=\frac{
\beta_h}{2}\left[\hbar\omega_1\coth\left(\beta_h\hbar\omega_1/2\right)-\hbar\omega_2\{(1-\mu)\coth\left(\beta_h\hbar\omega_2/2\right)+\mu
\coth\right(\beta_c\hbar\omega_2/2\left)\}\right].\nonumber
\end{align}
The appearance of the parameter $\mu$ appearing here may be viewed as a phenomenological way of incorporating 
non equilibrium effects arising from decoupling of the system from one bath and recoupling it to another.

In terms of  dimensionless quantities $a,b,c$ as
\begin{equation}
 \beta_c\hbar\omega_1=a,~~\frac{\omega_2}{\omega_1}=b, ~~\frac{\beta_h}{\beta_c}=c
\end{equation}
the expression above for the  efficienies $\eta_s^{\text{q}}$ and $\eta_s^{\text{cl}}$ in the 
classical and quantum read
\begin{align}
&\eta_{cl}(b,c)=\dfrac{1-c}{1+\mu\dfrac{(1-c)}{\ln b}},\\
&\eta_q(a,b,c)=\nonumber\\
&\dfrac{\ln\left(\dfrac{\sinh(abc/2)}
{\sinh(ac/2)}\right)-c\ln\left(\dfrac{\sinh(ab/2)}
{\sinh(a/2)}\right)}
{\ln\left(\dfrac{\sinh(abc/2)}
{\sinh(ac/2)}\right)
+\dfrac{ac}{2}\coth(ac/2)
-\dfrac{abc}{2}((1-\mu)\coth(abc/2)  -\mu\coth(ab/2))}.
\end{align}
In the experiments of Blickle and Bechinger \cite{7}
\begin{equation}
 a=9.50065\times 10^{-7},~~b=2.04922, c=0.845272.
\end{equation}
With $b$ fixed at this values we plot below the ratio $R=\eta_s^{\text{q}}/\eta_s^{\text{cl}}$ as 
a function of $a,c$ for two representative  values of $\mu$.
\vskip0.5cm
\noindent
\setlength{\unitlength}{1mm}
\begin{picture}(80,40)(0,0)
\put(5,-200){\epsfxsize=160mm\epsfbox{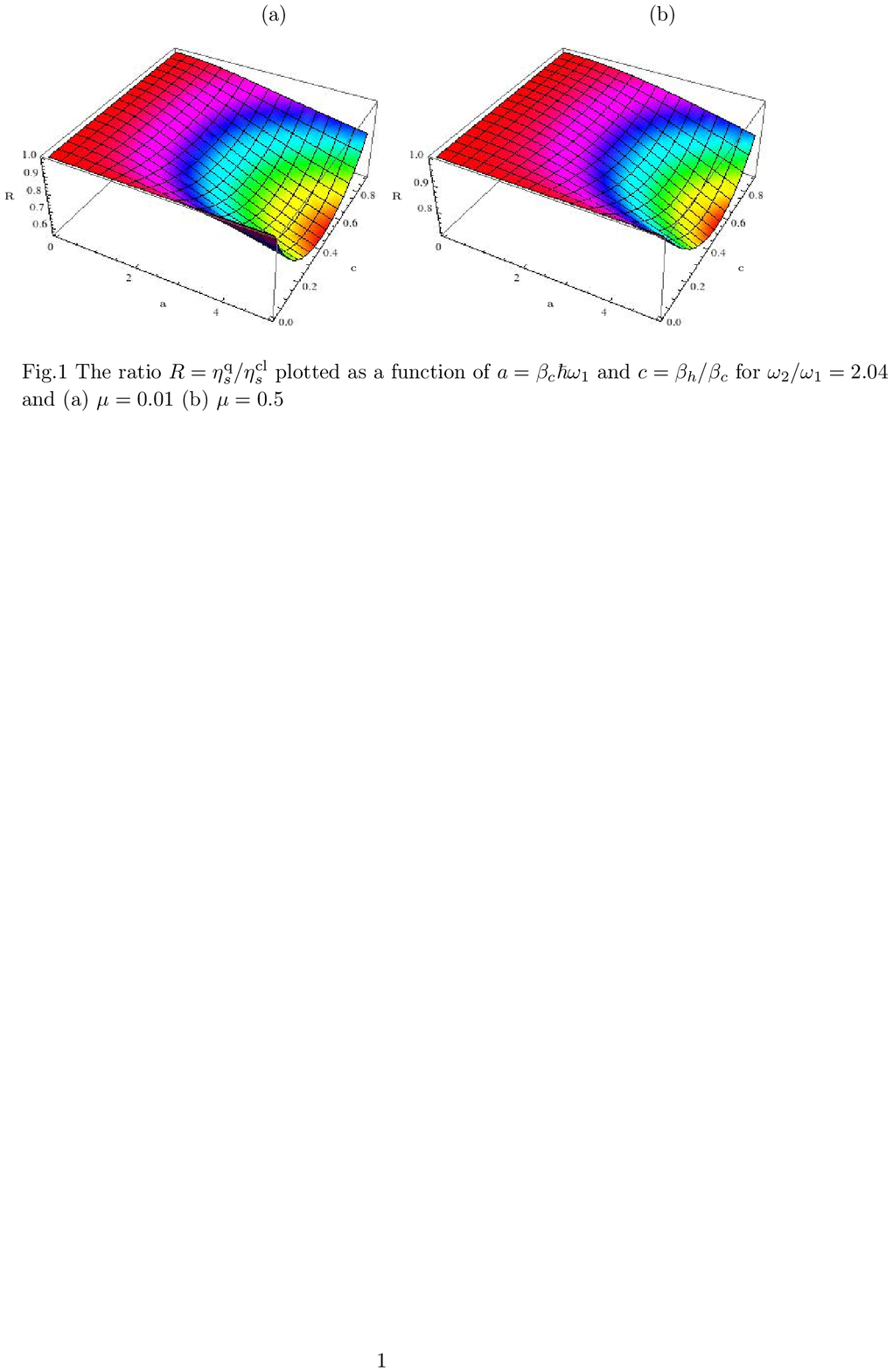}}
\end{picture}
\vskip2.5cm
\noindent

\section{Finite time corrections: Complementarity relations}
 We next consider the situation when the system starts out at equilibrium 
with a bath at temperature $T$, and the frequency is changed from its initial value 
$\omega_0$ to its final value $\omega_1$ in a finite time either isothermally ($T$ held fixed) or isentropically 
($\omega/ T$ held fixed) and focus on computing  finite time 
corrections to the standard thermodynamic results. Referring to $(\ref{1a})$-$(\ref{3a})$ we recall that 
while Stirling and Carnot engines involve the former operation, Carnot and Otto engines involve the latter. 
The scheme for computing finite time corrections developed below is similar in spirit to the 
adiabatic approximation in quantum mechanics and is a variant of the method 
formulated in \cite{18} in the context of the Fokker-Planck equation adapted  to the 
equations for the moments themselves.

The equations for the second moments that follow from the Langevin or the Fokker-Planck 
equation may be written as 
\begin{equation}
 \frac{d}{dt}X(t)= A(t)X(t)+Y(t),
\end{equation}
where 
\begin{equation}
 X(t)=\left(\begin{array}{c}<q^2>\\<qp>\\<p^2>\end{array}\right),
A(t)=\left(\begin{array}{ccc}0&\frac{2}{m}&0\\-m\omega^2(t)&-2\kappa&\frac{1}{m}
\\0&-2m\omega^2&-4\kappa\end{array}\right),
Y(t)=\left(\begin{array}{c}0\\0\\2D(t)\end{array}\right).
\end{equation}
(At this stage, as indicated, we allow  the frequency and the diffusion coefficients to be independent functions 
of $t$, Later however, we would specialise to situations appropriate to isothermal or isentropic variation 
of the frequency.)

Putting $t=s\tau$ and expanding $X(t)$ as 
\begin{equation}
 X(t)=X^{(0)}(s)+\frac{1}{\tau}X^{(1)}(s)+\cdots,
\end{equation}
we obtain
\begin{align}
 &A(s) X^{(0)}(s)+Y(s)=0 \Rightarrow X^{(0)}(s)=- A^{-1}(s)Y(s),\\
&X^{(1)}(s)=A^{-1}(s)\frac{d}{ds}X^{(0)}(s).
\end{align}
The first of these equations can be taken to describe the situation where the system is in the steady state 
corresponding to the instantaneous values of $\omega$ and $D$ and the second as describing deviations from 
this steady state. These equations then give 
\begin{equation}
 <q^2(s)>^{(0)}=\frac{D(s)}{2m^2\omega^2(s)\kappa}; <q(s)p(s)>^{(0)}=0; <p^2(s)>^{(0)}=\frac{D(s)}{2\kappa},
\label{78}
\end{equation}
and 
\begin{align}
& <q^2(s)>^{(1)}=-[\frac{8\kappa^2+2\omega^2(s)}{8\kappa\omega^2(s)}\frac{d}{ds}<q^2(s)>^{(0)}
\nonumber\\
&~~~~~~~~~~~~~~~~~+
\frac{1}{m\omega^2(s)}
\frac{d}{ds}<q(s)p(s)>^{(0)}+\frac{1}{4\kappa m^2\omega^2(s)}\frac{d}{ds}<p^2(s)>^{(0)}],\label{79}\\
&<q(s)p(s)>^{(1)}=\frac{m}{2}\frac{d}{ds}<q^2(s)>^{(0)}\\
&<p^2(s)>^{(1)}=-[\frac{m^2\omega^2(s)}{4\kappa}\frac{d}{ds}<q^2(s)>^{(0)}+\frac{1}{4\kappa}\frac{d}{ds}
<p^2(s)>^{(0)}].
\end{align}
These equations together with $(\ref{78})$ give  finite time corrections to the variances.  As the 
diffusion coefficient is a function of both $\omega$ and $T$ we now specialize to the situations where 
(a) $\omega$ is time dependent, $T$ is held fixed (Isothermal Case) (b) $\omega$,$T$  both are time dependent
but $\omega/T$ is held fixed (Isentropic case). With this in mind we may rewrite  
expression for $<q^2(s)>^{(1)}$ which we would need shortly as
\begin{align}
& <q^2(s)>^{(1)}\nonumber\\
&=\frac{\hbar}{4\kappa m\omega^2}\left[\left(\frac{4\kappa^2}{\omega^2}\right)\left(n(\omega,T)+\frac{1}{2}\right)\frac{d\omega}{ds}
 -\left(\frac{4\kappa^2}{\omega^2}+2\right)\omega\frac{d}{ds}\left(n(\omega,T)+\frac{1}{2}\right)
\right]. 
 \end{align}
In the isothermal case both the terms on the RHS contribute. On the other hand in the isentropic case only 
the first term contributes as during this process $\omega/T$ and hence $n(\omega,T)$ are held constant. 
In the following we confine ourselves to the isothermal case  and give the results for two physically interesting limiting cases corresponding to the overdamped  and weak dissipation dissipation regimes.

\begin{enumerate}
\item {\bf Overdamped regime}( $\kappa>>\omega$) 
\begin{align}
 &<q^2(s)>^{(1)}=\nonumber\\&~\frac{\kappa}{m\omega^5\beta}\left[\left(\frac{\beta\hbar\omega}{2}\right)\coth\left(\frac{\beta\hbar\omega}{2}\right)+\left(\frac{\beta\hbar\omega}{2}\right)^2
 {\text{cosech}^{2}} \left(\frac{\beta\hbar\omega}{2}\right)\right]\left(\frac{d\omega}{ds}\right).
\label{91a}
 \end{align}
\item {\bf Weak dissipation}( $\kappa<<\omega$) 
\begin{equation}
  <q^2(s)>^{(1)}=\frac{1}{2\kappa m\omega^3\beta}\left[\left(\frac{\beta\hbar\omega}{2}\right)
 {\text{cosech}} \left(\frac{\beta\hbar\omega}{2}\right)\right]^2\left(\frac{d\omega}{ds}\right).
 \label{92a}
\end{equation}
\end{enumerate}

We now compute expressions for the irreversible heat $Q_{\text{irr}}$ :
\begin{equation}
 Q_{\text{irr}}=F(i)-F(f)-\Delta W_{i\rightarrow f},
\label{95a}
\end{equation}
in an isothermal process at temperature $T$ from $i\rightarrow f$ arising from finite 
time corrections. Recalling that 
\begin{align}
&\Delta{ W}_{i\rightarrow f}
= -\int_{\omega_i}^{\omega_f} m\omega <q^2> d\omega \nonumber\\
&~~~~~~~~~~=-\int_{\omega_i}^{\omega_f} m\omega \left(<q^2>^{(0)}+\frac{1}{\tau}<q^2>^{(1)}\right) d\omega,
\end{align} 
and that 
\begin{equation}
  -\int_{\omega_i}^{\omega_f} m\omega <q^2>^{(0)} d\omega=F(i)-F(f),
\end{equation}
we have 
\begin{equation}
 Q_{\text{irr}}=\frac{1}{\tau}\int_{\omega_i}^{\omega_f} m\omega <q^2>^{(1)} d\omega \equiv T\frac{\Sigma}{\tau}
\end{equation}
From the way the quantity $\Sigma$ is defined above  it is clear that it would, in general, depend on both $T$ as well as 
on the manner in which $\omega$ is varied from its initial value $\omega_i$ to its final value $\omega_f$ in the time $\tau$. 
We now turn to the 
question as to what would be the minimum value of $Q_{\text{irr}}$ and hence that of $\Sigma$ in the weak dissipation 
and overdamped regimes. 

Using $(\ref{91a})$ and $(\ref{92a})$ and parametrizing $\omega (s)$  such that $\omega(0)=\omega_i,~ \omega(1)=
\omega_f$ we have 
\begin{equation}
Q_{\text{irr}}=\begin{array}{l}\dfrac{2\kappa}{\beta\tau}\int_{0}^{1}ds g(\omega (s))\left(\dfrac{d\omega}{ds}\right)^2~~~~~~~(\kappa >> \omega)\\
\\ \dfrac{1}{2\kappa \beta\tau}\int_{0}^{1}ds g(\omega(s))\left(\dfrac{d\omega}{ds}\right)^2
~~~~~(\kappa << \omega).
\end{array}
\end{equation} 
where
\begin{align}
g(\omega)=\begin{array}{l}\dfrac{1}{2\omega^4}\left[\left(\dfrac{\beta\hbar\omega}{2}\right)\coth\left(\dfrac{\beta\hbar\omega}{2}\right)+\left(\dfrac{\beta\hbar\omega}{2}\right)^2
 {\text{cosech}^{2}} \left(\dfrac{\beta\hbar\omega}{2}\right)\right]~(\kappa >> \omega)\\
\\
\dfrac{1}{\omega^2}\left[\left(\dfrac{\beta\hbar\omega}{2}\right)
{\text{cosech}} \left(\dfrac{\beta\hbar\omega}{2}\right)\right]^2
~~~(\kappa << \omega)\end{array}.
\end{align} 
The expression $Q_{\text{irr}}$ involve the functional 
\begin{equation}
 I[\omega]= \int_{0}^{1}ds g(\omega (s))\left(\frac{d\omega}{ds}\right)^2~.
\end{equation}
If we define $\Omega=f(\omega)$ and its inverse as $\omega=h(\Omega)$ and choose $f(\omega)$ to satisfy 
\begin{equation}
 \frac{df(\omega)}{d\omega}=\sqrt{g(\omega)},
\label{100a}
 \end{equation}
then we find that curve $\omega(s);~0\leq s \leq 1$ which minimizes  $I[\omega]$ is given by 
\begin{equation}
 \omega(s)=h(f(\omega(0))(1-s)+f((\omega(1))s)),
\end{equation}
and its minimum value of  by 
\begin{equation}
 I_{\text{min}}[\omega]=(f(\omega(1))-f(\omega(0)))^2.
\end{equation}

In the classical limit ($\beta\hbar\omega<<<1 $) in the strong damping regime $(\kappa >> \omega)$ we have 
\begin{equation}
 g(\omega)=\frac{1}{\omega^2} \Rightarrow f(\omega)=-\frac{1}{\omega};~h(\Omega)=-\frac{1}{\Omega},
\end{equation}
and hence 
\begin{equation}
 Q_{\text{irr}}^{\text{min}}=\frac{2\kappa K_BT}{\tau}\left[\frac{1}{\omega(1)}-\frac{1}{\omega(0)}\right]^2,
\end{equation}
in agreement with the results of Sekimoto and Sasa \cite{18} ( modulo an erroneous factor of 1/4 in the value of
$Q_{\text{irr}}^{\text{min}}$as quoted). This minimum value is realised along the curve
\begin{equation}
 \omega(s)=\left[\frac{s}{\omega(0)}+\frac{1-s}{\omega(1)}\right]^{-1}  ; 0\leq s\leq 1.
\end{equation}
Thus in the classical limit, in the overdamped regime, we obtain the following inequality for the 
product of the irreversible heat and the time taken to execute the step :
\begin{equation}
\tau\times Q_{\text{irr}}\geq 2\kappa K_BT\left[\frac{1}{\omega(1)}-\frac{1}{\omega(0)}\right]^2.
 \end{equation}
 Such  a relation is referred to in the literature as a thermodynamic complementarity relation, an analogue, both in 
 spirit and form, of the energy-time uncertainty relations in quantum mechanics. It should however be noted that the RHS of the above complementarity relation is independent of $\hbar$. 
 
 Again in the overdamped regime, but now in the low-temperature limit i.e $\beta\hbar\omega\rightarrow 0$, 
 we find that $g(\omega)\approx \beta\hbar/4\omega^3$ and the complementarity relation becomes 
 \begin{equation}
\tau\times Q_{\text{irr}}\geq \hbar(2\kappa)\left[\frac{1}{\sqrt{\omega(1)}}-\frac{1}{\sqrt{\omega(0)}}\right]^2.
 \end{equation}
 and one finds that $hbar$ now does appear on the RHS as one would expect in the limit of low temperature where quantum 
 effects become significant.
 
 Turning to the weak dissipation case, one finds that expression for $g(\omega)$ is such that the relevant integral 
 in $(\ref{100a})$ can be given in a closed form leading to the following complementarity relation: 
 \begin{equation} 
 \tau \times Q_{\text{irr}}\geq  \frac{K_BT}{2\kappa}
 \left[\log\left(\frac{\text{tanh}(\beta\hbar\omega(1)/4)}{\text{tanh}(\beta\hbar\omega(0)/4)}\right)\right]^2,
\end{equation}
valid for all values of $T$. In particular, in the classical limit it becomes :
\begin{equation}
 \tau \times Q_{\text{irr}}\geq \frac{K_BT}{2\kappa}\left[\log\left(\frac{\omega(0)}{\omega(1)}\right)\right]^2,
\end{equation}
and the curve $\omega(s),0\leq s\leq 1$ which minimizes $Q_{\text{irr}}$ now turns out to be 
\begin{equation}
 \omega(s)=\omega(0)^{(1-s)/2}\omega(1)^{s/2}.
\end{equation}

 We emphasise here that though we have presented explicit expressions for $Q_{\text{irr}}$ for  
 the weak and stong damping regimes, the results in $(\ref{78})$ and $(\ref{79})$ enable us to 
derive expressions for $Q_{\text{irr}}$ both for classical as well as quantum case 
without any specific assumptions on relative magintudes of $\kappa$ and $\omega$. Further, in the classical limit 
we find  that for an isothermal process from $i\rightarrow f$ carried out in a finite time $\tau$, 
$Q_{\text{irr}}$ has the structure 
\begin{equation}
 Q_{\text{irr}}= T\frac{\Sigma}{\tau},
\end{equation}
where $\Sigma$ is independent of $T$. On using the fact that for an isothermal process $F(i)-F(f)=T\Delta S_{i\rightarrow f}$, 
 we may rewrite  $(\ref{95a})$ as
\begin{equation}
 \Delta W_{i\rightarrow f}= T\left(\Delta S_{i\rightarrow f}-\frac{\Sigma_{i\rightarrow f}}
{\tau_{i\rightarrow f}}\right),
\label{109a}
\end{equation} 
This provides a convenient and physically useful way of parametrizing deviations from quasi-staticity in that 
in the limit $\tau_{i\rightarrow f}\rightarrow \infty$ one recovers the familiar results of equilibrium 
thermodynamics.

\section{Efficiency of the Stirling engine at maximum power}
In this section we would use the results of the previous section to analyse the efficiency of the Stirling 
engine at maximum power very much in the spirit of the earlier works in the context the Carnot cycle. 
We would closely follow the works of Schmiedl et al \cite{9} and of Esposito et al \cite{11}  
who analysed the question of the efficiency of the Carnot cycle at maximum power in the limit of low dissipation from 
fairly general considerations. In particular, in \cite{11}  it was shown that the Carnot efficiency at maximum power
$\eta_c^*$ is bounded below by $\eta_c/2$ and above by $\eta_c/(2-\eta_c)$ and that while the Curzon-Ahlborn efficiency \cite{19}
is reached in the limit of `symmetric' dissipation, the upper bound is realized in a completely asymmetric limit and 
coincides with the universal upper bound  derived in \cite{20} and \cite{21} from somewhat different considerations.  
 
Consider the situation when the isothermal steps $1\rightarrow 2$ and $3\rightarrow 4$ are carried out 
in finite times $\tau_h$ and $\tau_c$ respectively as indicated in $(\ref{1a})$. Power generated  during 
the Stirling cycle
is then
\begin{equation}
 P=\frac{\Delta W_{1\rightarrow 2}+\Delta W_{3\rightarrow 4}}{\tau_c+\tau_h}.
\label{110a}
\end{equation}
Also, as we have seen in Section VI, that the expression for the efficiency for the Stirling engine 
can be written as
\begin{equation}
 \eta_s^{\text{cl}}=\frac{\Delta W_{1\rightarrow 2}+\Delta W_{3\rightarrow 4}}
{\mu K_B(T_h-T_c)+\Delta W_{1\rightarrow 2}},
\label{111a}
\end{equation}
where $\mu\approx 0$ in the weak coupling regime and equal to $1/2$ in the overdamped regime. 

Using $(\ref{109a})$ and putting $\Sigma_{1\rightarrow 2}\equiv\Sigma_h,~\Sigma_{3\rightarrow 4}\equiv\Sigma_c,~
\Delta S_{1\rightarrow 2}=-\Delta S_{3\rightarrow 4}=\Delta S$ $(\ref{110a})$ and $(\ref{111a})$ become
\begin{align}
&P=\frac{(T_h-T_c)\Delta S-T_h\Sigma_h/\tau_h-T_c\Sigma_c/\tau_c}{\tau_c+\tau_h}\\
&\eta_s^{\text{cl}}=\frac{(T_h-T_c)\Delta S-T_h\Sigma_h/\tau_h-T_c\Sigma_c/\tau_c}
{\mu K_B(T_h-T_c)+T_h\Delta S-T_h\Sigma_h/\tau_h}.
\label{113a}
\end{align}
Maximizing $P$ with respect to $\tau_h$ and $\tau_c$ one finds that $P$ attains its maximum value for 
\begin{align}
 &\tau_h=\tau_h^*=2\frac{T_h\Sigma_h}{(T_h-T_c)\Delta S}\left(1+\sqrt{\frac{T_c\Sigma_c}{T_h\Sigma_h}}\right),\\
&\tau_c=\tau_c^*=2\frac{T_c\Sigma_c}{(T_h-T_c)\Delta S}\left(1+\sqrt{\frac{T_h\Sigma_h}{T_c\Sigma_c}}\right).
\end{align}
Substituting these values for $\tau_h$ and $\tau_c$ in $(\ref{113a})$ one finds that the efficiency for the 
Stirling engine at maximum power is given by
\begin{equation}
 \eta_s^{\text{cl}*}=\frac{\eta_c\left(1+\sqrt{\frac{T_c\Sigma_c}{T_h\Sigma_h}}\right)}{
\left(1+\sqrt{\frac{T_c\Sigma_c}{T_h\Sigma_h}}\right)^2+\frac{T_c}{T_h}\left(1-\frac{\Sigma_c}{\Sigma_h}\right)
+\frac{2\mu\eta_c}{\log\left(\frac{\omega_2}{\omega_1}\right)}}.
\end{equation}
We now consider two cases:
\vskip4mm 
\noindent
{\bf Case A $\mu=0$}
\vskip4mm
\noindent
In the extreme weak dissipation regime i.e. $\mu=0$, one recovers  results similar to those 
in \cite{9},\cite{11} in the context of the Carnot cycle :
\begin{enumerate}
 \item In the symmetric case i.e $\Sigma_c/\Sigma_h=1$, $\eta_s^{\text{cl}*}$ equals the Curzon-Ahlborn 
efficiency $ \eta_{CA} =1-\sqrt{T_c/T_h}$:
\begin{equation}
 \frac{\Sigma_c}{\Sigma_h}=1:~~~~\eta_s^{\text{cl}*}=\eta_{CA}.
\end{equation}
\item $\eta_s^{\text{cl}*}$ is bounded by $\eta_c/2$ and $\eta_c/(2-\eta_c)$
\begin{equation}
\eta_c/2\leq \eta_s^{\text{cl}*}\leq \eta_c/(2-\eta_c).
\end{equation}
The upper and the lower bounds respectively correspond to $\Sigma_c/\Sigma_h \rightarrow 0$ and 
$\Sigma_c/\Sigma_h \rightarrow \infty$
\end{enumerate}
\vskip4mm 
\noindent
{\bf Case B $\mu\neq 0$}
\vskip4mm
\noindent
For small but non zero $\mu<\frac{1}{2}\log(\omega_2/ \omega_1)$ these results get modified to those given below
\begin{enumerate}
 \item In the symmetric case i.e $\Sigma_c/\Sigma_h=1$, $\eta_s^{\text{cl}*}$ is less than the Curzon-Ahlborn 
efficiency \cite{19} $ \eta_{CA} =1-\sqrt{T_c/T_h}$:
\begin{equation}
 \frac{\Sigma_c}{\Sigma_h}=1:~~~~\eta_s^{\text{cl}*}=\frac{\eta_{CA}}{1+\left(\frac{\mu}{\log\left(\omega_2/
\omega_1\right)}\right)\left(\frac{2\eta_{CA}}{2-\eta_{CA}}\right)}<~\eta_{CA}.
\end{equation}
\item $\eta_s^{\text{cl}*}$ is bounded by $\eta_c/2$ and $\eta_s/(2-\eta_s)$
\begin{equation}
\eta_c/2\leq \eta_s^{\text{cl}*}\leq \eta_s/(2-\eta_s).
\end{equation}
As before, the upper and the lower bounds respectively correspond to $\Sigma_c/\Sigma_h \rightarrow 0$ and 
$\Sigma_c/\Sigma_h \rightarrow \infty$
\end{enumerate}
On the other hand if $\mu>\frac{1}{2}\log(\omega_2/ \omega_1)$, one finds that 
\begin{equation}
\eta_s^{\text{cl}*}\leq \eta_c/2.
\end{equation}
In the figures below we display the bounds on $\eta_s^{\text{cl}*}$ for $\mu=0.001,0.1,0.2,0.4$ with 
$\omega_2/\omega_1$ taken to be $2.05$ where we also give the plots for $\eta_c,\eta_{CA}$ and $\eta_{c}/2$  
for comparison.
\vskip1cm
\noindent
\setlength{\unitlength}{1mm}
\begin{picture}(80,80)(0,0)
\put(5,-180){\epsfxsize=160mm\epsfbox{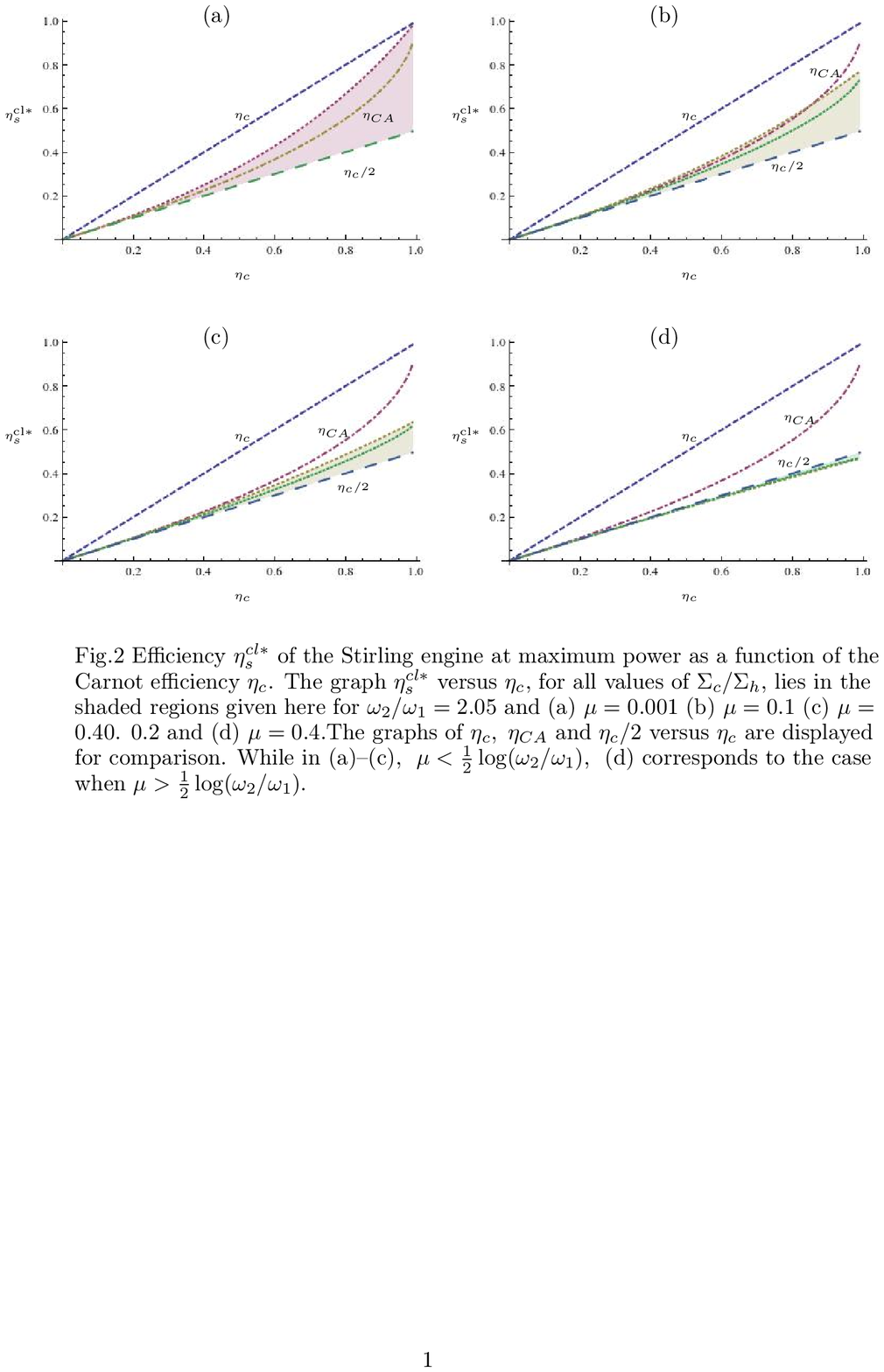}}
\end{picture}
\vskip8cm
\section{Conclusions}
In this work we have developed a microscopic framework for computing efficiencies of 
quantum/classical brownian motors realized by a harmonic oscillator. Two exactly solvable 
models for frequency modulation are presented. In the context of the Stirling Engine we have analytically 
treated the question of coupling the system at equilibrium with a bath at low temperature 
to a bath at higher temperature and the role that various time scales play in this process 
and have shown that these have strong influence on its efficiency. Further we have developed a 
procedure for computing finite time corrections to the quantitities of interest needed for calculating 
the efficiency of the the three engines considered - Stirling, Carnot and Otto, and have derived the 
thermodynamic complementarity relations in the overdamped and underdamped situations both in the high 
as well as low temperature limits. In the spirit 
of the works reported in \cite{9} and \cite{11} on the Carnot engine, we have analysed in detail the efficiency of the Stirling engine at maximum power and have investigated the role of dissipation parameters thereon. 
Though in this work we have exclusively considered interactions between the system and a thermal bath, the formalism can easily be extended to  situations where the thermal bath is replaced by a squeezed thermal bath 
bringing with it new parameters and thereby ushering in new possibilities that have no classical 
analogues. 

In the present work we have modelled the three heat engines after the quantum harmonic oscillator. It is of interest to 
carry out a similar analysis for finite state quantum systems. Indeed the entire perspective on heat pumps, refrigeratators and 
heat engines developed in \cite{3} is based on the simplest of quantum systems -- a qubit. Though in that context a convenient 
Fokker-Planck framework would no longer be available  we expect that the methodology developed here applied directly to the 
master equation would be useful there as well. We hope to return to this and related questions in the near future. 
\vskip1.5cm
\noindent
{\bf Acknowledgment:} One of (SC)  gratefully acknowledges the
hospitality extended to him by the Department of Physics, Oklahoma State University, Stillwater where a large part of this 
work was carried out.

                                                          \

\end{document}